\documentclass[prl,twocolumn,aps,superscriptaddress,showpacs,floatfix]{revtex4}

\usepackage{longtable}
\usepackage{graphicx}
\usepackage{dcolumn}
\usepackage{bm}
\begin{document}
\title{Massless Dirac fermions in a laser field as a counterpart of graphene superlattices.}

\author{Sergey E. Savel'ev}
\affiliation{Department of Physics, Loughborough University,
Loughborough LE11 3TU, United Kingdom}
\author{Alexandre S. Alexandrov}
\affiliation{Department of Physics, Loughborough University,
Loughborough LE11 3TU, United Kingdom}

\begin{abstract}
We establish an analogy between spectra of Dirac fermions in laser fields and an electron spectrum of graphene superlattices 
formed by static 1D periodic potentials.
The general relations between a laser-controlled spectrum where electron momentum
depends on the quasi-energy and a superlattice mini-band spectrum in graphene are derived. 
%
%
As an example we consider
two spectra generated by a pulsed laser and by a step-like electrostatic potential. We also calculate the graphene excitation spectrum
in continuous strong laser fields in the resonance approximation for linear and circular polarizations and show that
circular polarized laser fields cannot be reduced to any graphene electrostatic superlattice. Some physical phenomena related to the peculiar 
graphene energy spectrum in the strong electromagnetic field are discussed.
  
\end{abstract}
\pacs{05.40.-a, 05.60.-k, 68.43.Mn} \maketitle

A huge surge of interest to graphene (see e.g., \cite{Novoselov05,review}) as the only two dimensional material known so far 
stimulates studies of one dimensional and two dimensional graphene superlattices (see e.g., \cite{super}). In analogy to usual semiconducting superlattices, it is commonly accepted that such graphene superlattices should allow to manipulate the electron spectra and transport properties
in graphene-based electronics. However, there is still a limited number of studies on how time-dependent 
electric field (for instance laser field) can affect
both the electron spectrum and transport in graphene. Most studies done so far are focused on a perturbative response of graphene in weak (probe)
electromagnetic time-dependent fields \cite{lowfields}. A growing experimental demand \cite{demand} on the data analysis of optical properties and electron transport in graphene in strong laser fields is the main motivation of this work.

Another motivation is to study temporal-spatial symmetry for massless Dirac fermions in periodic time or 
spatial electric fields. Spatial periodic potentials $U(x)$ with its period much larger than the interatomic distance transforms the Dirac cone energy spectrum $\varepsilon(p_x,p_y)$ with electron momentum ${\vec p}=(p_x,p_y)$ into a set of minibands $\varepsilon_n(p_x,p_y)$ \cite{super} where the quasi-momentum $p_x=\hbar k_x$ or the electron wave vector $k_x$ takes values within the first superlattice Brillouin zone $-\pi/L<k_x<\pi/L$, while
the y-momentum $p_y=\hbar k_y$ is unbound. 
Here $L$ is the spatial period of an electric field $E_x(x)$ and an integer $n$ numerates different subbands (which are usually separated by gaps). 
In contrast to the above picture, applying time periodic electric fields $E_x(t)$ should change the spectrum of electrons in accordance with the
Floquet theory stating that the electron energy should become a quasi-energy 
$\varepsilon$ bounded within its Brillouin zone $-\pi/T<{\cal E}<\pi/T$ with ${\cal E}
=\varepsilon/\hbar$ and $T$ is the temporal period of homogeneous electric field $E_x(t)$ applied along the x-axis. Therefore, the electron spectrum in time-periodic laser fields could be written in the form $k_x=k_x(\varepsilon,k_y)$. This analogy raises several questions, including 
the possibility of gaps in the spectrum for either momentum $k_x$ or quasi-energy $\varepsilon$. 

In this context, we would like to note that a study of conventional gapped semiconductors in strong laser fields has a 
long history. It was shown
that beyond linear response theory, a strong harmonic field opens energy gaps caused by the high-frequency Stark effect
(e.g., \cite{gal}). The location of the gaps in momentum space and their size can be controlled independently by changing 
either frequency or amplitude of laser fields.

In this letter we show that two problems of electron excitation spectra in graphene in either (i) spatial or (ii) temporal periodic potential have a
deep analogy allowing to obtain an unknown spectrum if the dual spectrum is calculated or measured. 
This opens a possibility to indirectly measure the spectrum of complicated graphene superlattices with no expensive and 
time-consuming nanofabrication techniques needed to generate a desirable spatial variation of electric fields 
on a nano-scale but rather applying easily adjustable temporally periodic laser electric fields. Then, we
consider two specific examples: (1) an analytically solvable model for periodic piece-wise constant (square wave-like) potentials, which can be realized by using a sequence of laser pulses or by making spatial-potential steps $\pm U_0$, (2) monochromatic linearly polarized laser fields ($E_x(t)\propto \sin \omega t$), which can be analyzed by using the resonance approximation (e.g., \cite{gal}). 
As an example of a laser field problem irreducible to any spatially periodic electrostatic 
graphene superlattices we consider circularly polarized electric fields generating a quasi-isotropic 
gap in the energy spectrum in contrast to an anisotropic gap with gapless points for linearly polarized laser fields. 

{\it Spatial-temporal duality.---\/} Let us consider two related problems of graphene subject to either spatially periodic $E_x(x)=-dU/dx$ or time periodic $E_x(t)=-dA_x/dt$ electric fields. Here we introduce the static electric potential $U(x)$ and time-dependent vector potential $A_x(t)$  which are periodic functions of their arguments, {\it i.e.}, $U(x+L)=U(x)$ and $A_x(t+T)=A_x(t)$. In the low energy limit, the behaviour of the charge carriers can be described by the 2D Dirac equation:
$v_F({\vec\sigma}({\vec p}-{\vec A}))\Psi + U\Psi = i\hbar \partial \Psi/\partial t$ with a two component wave function $\Psi=(\Psi_A, \Psi_B)$ corresponding to electrons in two triangular graphene sublattices and $v_F$ is the Fermi velocity. For the case of spatially periodic electric fields (graphene superlattices), {\it i.e.,} $A_x=0$, the wave function can be written as follows $\Psi_{A,B}=\varphi_{A,B} (x)\exp(-i{\cal E} t+ik_yy)$ and the coupled equations for $\varphi_{A,B}$
have the form:
\begin{equation}
i\frac{d}{dx} \left( \begin{array}{c} \varphi_A \\ \varphi_B \end{array} \right) = 
\left(\begin{array}{cc} ik_y & -{\cal E}+U(x) \\ -{\cal E}+U(x) & -ik_y \end{array}\right) \left( \begin{array}{c} \varphi_A \\ \varphi_B 
\end{array} \right). \label{sup} 
\end{equation}
Here and below we use $\hbar=v_F=1$ and measure  wave vectors in the unit reciprocal lattice  vector of the superlattice, $\pi/L$, or of the graphene lattice, $\pi/a$.
For the case of time dependent electric fields $E(t)=-dA_x/dt$ and $U(x)=0$, the solution can be written as $\Psi_{A,B}=\phi_{A,B}\exp(i k_xx+ik_yy)$, where
\begin{equation}
i\frac{d}{dt} \left( \begin{array}{c} \phi_A \\ \phi_B \end{array} \right) = 
\left(\begin{array}{cc} 0 & k_x-A_x(t)-ik_y \\ k_x-A_x(t)+ik_y & 0 \end{array}\right) \left( \begin{array}{c} \phi_A \\ \phi_B 
\end{array} \right). \label{time_sup}
\end{equation}                      
Both sets (\ref{sup}) and (\ref{time_sup}) 
of differential equations belong to the same class of ordinary differential equations with periodic coefficients so that one can apply the 
Floquet (or Bloch) theory to classify all possible states. To see a deeper analogy we can use new variables ${\tilde \phi}_A=(\phi_A-i\phi_B)/\sqrt{2}$
and ${\tilde \phi}_B=(\phi_B-i\phi_A)/\sqrt{2}$ for time-dependent potentials, so that both sets (\ref{sup}, \ref{time_sup})
can be written in the same form
\begin{equation}
i\frac{d}{d\xi} \left( \begin{array}{c} f_A \\ f_B \end{array} \right) = 
\left(\begin{array}{cc} i\beta & Q+\mu(\xi) \\ Q+\mu(\xi) & -i\beta \end{array}\right) \left( \begin{array}{c} f_A \\ f_B 
\end{array} \right). \label{Floquet} 
\end{equation}
where $\mu(\xi+{\cal L})=\mu(\xi)$ is a periodic function of $\xi$ with the period ${\cal L}$.
Here, $\xi=x$, $\mu=U$, $Q=-{\cal E}$, $\beta=k_y$, and ${\cal L}=L$ for spatial-dependent electric fields while $\xi=t$, $\mu=-A_x$, $Q=k_x$,  
$\beta=-ik_y$, and ${\cal L}=T$ for time-dependent homogeneous electric field. All the solutions 
of equations (\ref{Floquet}) can be classified using Floquet exponents
$f_{A,B}(\xi+{\cal L})=\exp(iK{\cal L})f_{A,B}(\xi)$ with Floquet multipliers $-\pi/{\cal L}<K<\pi/{\cal L}$. Spectrum linking Floquet
multiplier (quasi-momentum $k_x$ for spatially varying potentials and quasi-energy ${\cal E}$ for time-dependent laser fields) depends on
the shape of the periodic potential $\mu$, but, in general, can be written as $F(K,Q,\beta)=0$. Therefore, the spectrum of Dirac electrons in the
periodic superlattices with $U=P(x)$ and the spectrum of Dirac electrons in laser fields with $A_x=-P(t)$ described by the same function $P$ are defined by the same spectral equation, i.e., $F(k_x,-{\cal E},k_y)=0$ for spatial oscillating potentials and $F(-{\cal E},k_x,ik_y)=0$ for laser fields.
For instance, if we measure a quasi-energy spectrum for a laser field having a certain time-profile $P(t)$,
 then we can easily obtain the spectrum of graphene superlattices with a spatial potential of the same shape $U(x)=-P(x)$.  

\begin{figure} \vspace{-0.7cm}
\includegraphics[width=7cm]{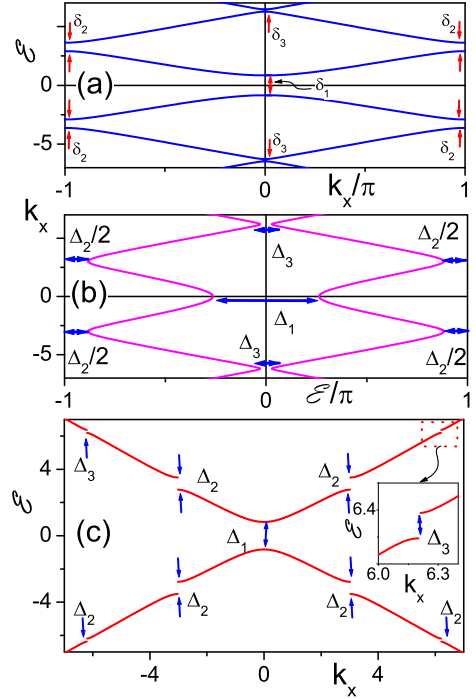}
\vspace{-0.7cm}
\caption{(Color online) (a) Mini-band electron structure for a graphene sample in the step-like spatial potential $U(x)=U_0 {\rm sign}(\sin(2\pi x/L))$
with $U_0=1,\ L=1, k_y=1$. The gaps ($\delta_1, \delta_2, \delta_3, ...$) between different zones are clearly seen. 
(b) Reduced zone momentum-quasi-energy spectrum $k_x({\cal E})$
for graphene in the laser field $A=A_0 {\rm sign} (\sin (2\pi t/T))$ with $A_0=1, T=1, k_y=1$. There are no gaps for momentum but
not all values of quasi-energy can be reached resulting in gaps in quasi-energy $\Delta_1, \Delta_3, \Delta _3$ marked on the figure. The physical meaning of the energy gaps are readily seen in the extended zone representation (c).} \label{fig1}
\end{figure}

{\it Periodic spatial step-like potential versus pulsating laser fields.---\/} Generally it is not possible to
write down the solution of Eqs.~(\ref{Floquet}) for $k_y\ne 0$ and, thus, to derive the spectral equation $F(K,Q,\beta)=0$ for an arbitrary 
function $\mu(\xi)$. However, there are several
particular cases when we can obtain analytical solutions and, thus, the corresponding spectra of Eqs. (\ref{Floquet}). One interesting example is the step-like spatial
periodic function $U(x)=U_0{\rm sign}[\sin (2\pi x/L)]$ and the dual periodic temporal potential $A_x=A_0{\rm sign}[\sin(2\pi t/T)]$, both resulting in $\mu(\xi)=\mu_0{\rm sign}[\sin (2\pi \xi/{\cal L})]$ (with $\mu_0=U_0$ or $\mu_0=A_0$). Here ${\rm sign(x)}=1$ for $x>0$ and $-1$ for $x<0$. A 
solution for $f_{A,B}$ can be written as $f_{A,B}=C_{A,B; 1}^{(+)}
\exp(i\kappa_1\xi)+C_{A,B; 1}^{(-)}\exp(-i\kappa_1)\xi$ with $C_{B; 1}^{(\pm)}=(\mp \kappa_1-i\beta)C_{A;1}^{(\pm)}/(Q-\mu_0)$ for $-{\cal L}/2<\xi<0$
and $f_{A,B; 2}=C_{A,B; 2}^{(+)}\exp(i\kappa_2\xi)+C_{A,B; 2}^{(-)}\exp(-i\kappa_2)\xi$ with $C_{B; 2}^{(\pm)}=(\mp \kappa_1-i\beta)C_{A;2}^{(\pm)}/(Q+\mu_0)$ for $0<\xi<{\cal L}/2$. Here we introduce $\kappa_{1,2}$ as
$\kappa_{1,2}=\sqrt{(Q\pm \mu_0)^2-\beta^2}$
where the sign ''$-$'' should be taken for subindex 1 and ''$+$'' for subindex 2, while subindexes 1 and 2 refer to the 
regions $-{\cal L}/2<\xi<0$ and
$0<\xi<{\cal L}/2$ respectively. It is important to stress here that $\kappa_{1,2}$ can become imaginary for
spatially oscillating fields when $\beta^2=k_y^2>(Q\pm E_0)^2$, but $\kappa_{1,2}$ are always real for time-oscillating fields since
$\beta^2=(ik_y)^2=-k_y^2$. Using both, the matching conditions for $f_{A,B; 1,2}$ at $\xi=0$ and the quasiperiodic conditions $f_{A,B}(\xi={\cal L}/2)=\exp(iK{\cal L})
f_{A,B}(\xi=-{\cal L}/2)$ we can derive four homogeneous equations for four variables $C_{A; 1,2}^{\pm}$ which can be solved only if its determinant
is equal to zero, which defines the spectral equation:
\begin{eqnarray}
\cos\left(K{\cal L}\right)&=&\cos\left(\frac{\kappa_1{\cal L}}{2}\right)\cos\left(\frac{\kappa_2{\cal L}}{2}\right) \label{spectrum} \\&-&\frac{Q^2-\mu_0^2-\beta^2}{\kappa_1\kappa_2}\sin\left(\frac{\kappa_1{\cal L}}{2}\right)\sin\left(\frac{\kappa_2{\cal L}}{2}\right)\nonumber
\end{eqnarray}        
Following our duality rule we have to substitute $K=k_x, Q=-{\cal E}, \beta=k_y$ and ${\cal L}=L$ for graphene superlattices ($E(x)$-electric fields)
and $K=-{\cal E}, Q=k_x, \beta=ik_y$ and ${\cal L}=T$ for time-dependent laser pulse fields. As mentioned above, $\kappa_{1,2}$ can become imaginary
for spatial oscillating fields, which results in the right-hand-side of equation (\ref{spectrum}) to exceed 1, thus, giving no roots to this equation
for such values of energy (energy gaps $\delta_1,\delta_2,..$, see Fig. 1a). In contrast, the right-hand side of equation (\ref{spectrum}) never exceeds 1 (moreover is always less than 1 for $\beta^2=-k_y^2<0$), thus, the equation has solutions for any $Q=k_x$ and $k_y$ producing no gaps in the momentum. However, some of the values of quasi-energy ${\cal E}=-K$ near the boundaries $\pm T/\pi$ of the Brillouin zone cannot be reached resulting in gaps in quasi-energy (see Fig. 1b). Note that
the energy or quasi-energy spectrum of equation (\ref{spectrum}) has no gaps for either spatial or temporal oscillating electric fields if $k_y=0$.
This last property could be obtained for any spatial $U(x)$ and temporal $A_x(t)$ dependence of electric fields, which can be proven by using
general solution (for $k_y=0$) obtained in \cite{ishikawa} for the Dirac equitation (\ref{time_sup}). 
Sometimes it is more convenient to replace the quasi-energy spectrum (Fig. 1)
in the reduced zone representation $-T/\pi<\varepsilon\leq T/\pi$ by the spectrum in the extended zone representation shown in Fig. 1c where energy gaps $\Delta_1, \Delta_2,...$ are clearly seen.

{\it Harmonic linearly polarized laser field.---\/} Unfortunately, an exact solution for the harmonic potential $A_x(t)=A_0\cos(\omega t)$
with frequency $\omega=2\pi/T$ or its dual spatially periodic 
potential $U(x)=U_0\cos(2\pi x/L)$ is unknown. However, we can use the so-called resonance approximation to 
calculate the spectrum of Eqs. (\ref{time_sup}). Namely, we can seek a solution in the form $\phi_{A,B}=\exp(i{\cal E}t)[D_{A,B}^{(+)}\exp(i\omega t/2)+D_{A,B}^{(-)}\exp(-i\omega t/2)]$. When ignoring all fast oscillating terms [i.e., $\exp(\pm 3i\omega t/2), \exp(\pm 5i\omega t/2), ...$], 
we can obtain a closed set of four homogeneous equations for four variables $D_{A,B}^{(\pm)}$, which can be solved only if its 
determinant is equal to zero resulting in the 
spectrum:
\begin{equation}
{\cal E}=\pm\left(\frac{\omega^2+A_0^2}{4}+(k_x^2+k_y^2)\pm\sqrt{(k_x^2+k_y^2)\omega^2+A_0^2k_x^2}\right)^{1/2}
\end{equation}
where $\pm$ in front and inside the brackets should be taken independently, producing four independent branches shown in Fig. 2b,c. 

In agreement with our general statement, the oscillating laser field opens a gap in the spectrum, which can be controlled 
by both the frequency $\omega$ and the field intensity $A_0$. The gap between different branches 
of the spectrum drops to zero for $k_y=0$ and any field intensity $A_0$(as expected from our general consideration) (Fig. 2b), while gaps
are seen for nonzero $k_y$ (Fig. 2c). Physical meaning of four different branches can be easily understood in the limit of $A_0\rightarrow 0$. Indeed, 
in addition to the usual energy cone (Fig. 2a(left); also, the dark-blue line for electrons and the dark-green line for holes in Fig. 2a(right)), 
we have to consider the energy spectrum of a hole and a photon (a cone shifted up
by $\omega$ shown in Fig 2a(right) in light-green) and the spectrum of an electron and an emitted photon (a cone 
shifted down by $\omega$ shown in Fig. 2a(right) in light-blue). Gaps can be formed at the crossing points of these spectra for 
nonzero field intensity (so-called the high-frequency Stark effect). In this case, we recover the spectrum shown on Fig. 2b for the energies 
${\cal E}\pm\omega/2$ since the wave function contains the terms $D_{A,B^{(\pm)}}\exp(i({\cal E}\pm \omega/2)t)$.

\begin{figure}
\vspace{-0.7cm}
\includegraphics[width=7cm]{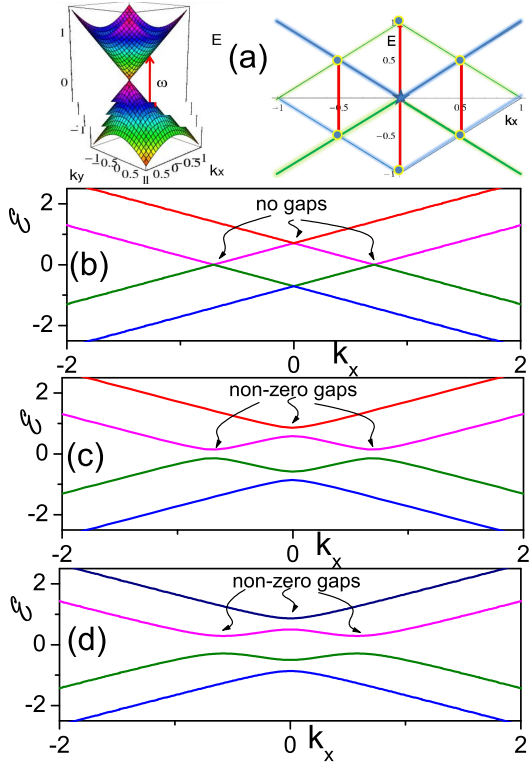}
\vspace{-0.3cm}
\caption{(Color online) (a, left) The sketch of the excitation of the electron Dirac cone in a harmonic laser field with the frequency $\omega$,
arrow shows electron transitions under the electric field. (a, right) Crossing of electron-photon terms under the action of the electric field: the Dirac cone shifts both up (representing the hole and photon state) and down (representing an electron which has already emitted a photon), the crossing 
of all these three cones [the original Dirac cone (in dark-green and dark-blue) and 
the two cones shifted by $\omega$ (in light-green and light-blue)] results in the opening of gaps between the
corresponding energy bands. (b) Electron spectrum ${\cal E}(k_x)$ for graphene in a linearly polarized electric laser field with $\omega=1, A_0=1, k_y=0$ (all measured in $\pi/T$): there are no gaps between the energy zones in agreement with our spatial-temporal symmetry conclusions. (b) The same as (a) but for nonzero y-momentum $k_y=1$, resulting in the opening of gaps between the energy bands. 
(c) The same as (a) with zero y-momentum, but for the circular polarized
laser field $\alpha=\pi/2$ also resulting in the opening of the gaps 
between zones since there is no analogy between the circular polarized light and any
periodic electrostatic potential.} \label{fig2}
\end{figure}

{\it Circular laser fields.---\/} An interesting example of a problem, when a time-dependent laser field has no analogue in a spatially periodic electrostatic field, is the circular polarized harmonic laser field described by a vector
potential ${\vec A}=(A_x(t), A_y(t))$ with $A_x(t)=(A_0/\sqrt{2})\cos \omega t$ and $A_y(t)=(A_0/\sqrt{2})\cos(\omega t+\alpha)$. Seeking a wave function in the form $\Psi_{A,B}=\phi_{A,B}\exp(ik_xx+ik_yy)$ we recover equation (\ref{time_sup}) with replacement $k_y$ by $k_y-A_y(t)$. Using $\phi_{A,B}=\exp(i{\cal E}t)[D_{A,B}^{(+)}\exp(i\omega t/2)+D_{A,B}^{(-)}\exp(-i\omega t/2)]$ and ignoring higher harmonics in the resonance approximation, we again obtain the
set of four homogeneous equations for $D_{A,B}^{(\pm)}$ with the spectrum
\begin{eqnarray}
&{\cal E}&=\pm \Biggl[\frac{\omega^2+A_0^2}{4}+(k_x^2+k_y^2)\nonumber\\
&\pm&\sqrt{(k_x^2+k_y^2)(\omega^2+\frac{A_0^2}{2})+\frac{A_0^4}{16}\sin^2\alpha+A_0^2k_xk_y\cos\alpha}\;\Biggr]^{1/2}
\end{eqnarray} 
Note that the last spectrum has no more requirements of zero gaps for $k_y=0$, so the four quasi-energy branches can be well separated
for a properly chosen phase shift $\alpha$ (see Fig. 3). Thus, the spectrum becomes even more tuneable by changing
frequency, amplitude and/or phase of the circular polarized laser field.     

\begin{figure}
\includegraphics[width=8.5cm]{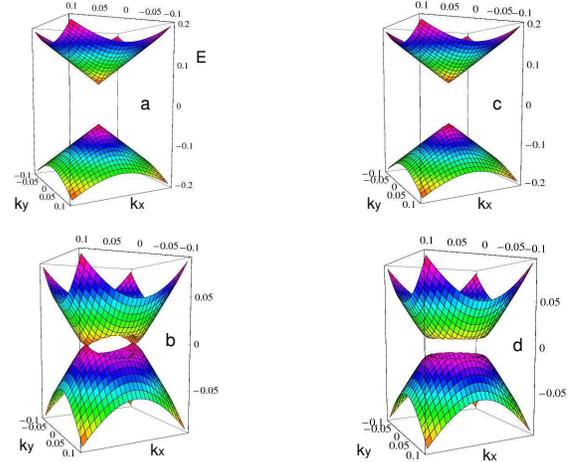} \vspace{-0.6cm}
\caption{(Color online) Electron spectra ${\cal E}(k_x,k_y)$ for graphene in linear polarized 
(a[upper and bottom energy branches],b[two middle energy branches]) 
and circular polarized (c[upper and bottom energy branches],d[two middle energy branches]) laser fields with
parameters $A_0=0.03, \omega=0.1$ (measured in units $\hbar v_F \pi/a$ with $a$ being the atomic lattice constant). Two points for $k_y=0$ where two middle energy branches touch one another
are clearly seen for linearly polarized electromagnetic fields, while these branches are well separated for any $k_x, k_y$
for the circular polarized electromagnetic fields ($\alpha=\pi/2$) (c,d).} \label{fig3}
\end{figure}

{\it Perspectives and conclusions.---\/} The optical gap in the spectrum of graphene electrons generated by the electromagnetic field can be responsible for unusual electron dynamics and kinetics, for instance, the pseudo-periodic ${\cal E}(k_x(t))$ Bloch oscillations can be induced by applying a small DC electric field. Stationary distributions of electrons in strong laser fields can be very peculiar compared to the electron distribution \cite{gal} in conventional (gapped) semiconductors. 
In the latter case with a large band-gap one can readily reach a quasi-equilibrium Fermi-Dirac distribution of electrons pumped  into  the 
conduction band and holes in the valence band, and an insulating state due to the isotropic optical gap
(the so-called ``optical'' insulator \cite{asael}). However, such degenerate distributions seem to be improbable in graphene, since 
photoelectrons can recombine with photoholes with about the same rate as their energy relaxation rate. 
Moreover, the spectrum predicted here can be directly measured by applying a weak probe field in additional to strong time-periodic laser or
spatially periodic electrostatic fields. The duality between the electron spectrum in spatially periodic electrostatic fields and 
time-periodic laser fields provides an experimental access to the complex spatially periodic graphene superlattices that was impossible before
due to the expensive, hardly controlled and not easily tuneable nano-fabrication procedure. 
 
We thank Prof. C.M. Linton and Dr. V. Zalipaev for discussions. SES acknowledges support from the EPSRC (EP/D072581/1).

\end{document}